\documentclass[notitlepage,a4paper,aps,prd,onecolumn,superscriptaddress,nofootinbib,groupedaddress]{revtex4-2}

\usepackage{amsmath}
\usepackage{amsfonts,color}
\usepackage{amsthm}
\usepackage{ulem}
\usepackage{cancel}
\usepackage{empheq}
\usepackage{amssymb,float}
\usepackage{booktabs,multirow}
\usepackage{titlesec}
\usepackage{hyphenat}
\usepackage{ragged2e}
\usepackage{amsmath}
\usepackage{accents}
\usepackage[utf8]{inputenc}
\usepackage{color}
\usepackage{hyperref}
\usepackage{enumitem}
\usepackage{tikz}
\usetikzlibrary{shapes.geometric}
\usetikzlibrary{arrows.meta,arrows}

\titleformat{\paragraph}
{\normalfont\normalsize\bfseries}{\theparagraph}{1em}{}
\titlespacing*{\paragraph}
{0pt}{3.25ex plus 1ex minus .2ex}{1.5ex plus .2ex}

\allowdisplaybreaks
\hypersetup{
    colorlinks=true,
    linkcolor=blue,
    filecolor=magenta,      
    citecolor=blue
}

\begin{document}

\title{Degrees of Freedom of New General Relativity:\\ Type 4, Type 7, and Type 9}
\author{Kyosuke Tomonari
\footnote{Current affiliation: Institute of Astrophysics, Central China Normal University, Wuhan 430079, China}
}
\email{ktomonari.phys@gmail.com}
\affiliation{Department of Physics, Institute of Science Tokyo, 2-12-1 Ookayama, Meguro-ku, Tokyo 152-8551, Japan\\}
\affiliation{Interfaculty Initiative in Information Studies, Graduate School of Interdisciplinary Information Studies, The University of Tokyo, 7-3-1 Hongo, Bunkyo-ku, Tokyo 113-0033, Japan\\}

\begin{abstract}
We investigate degrees of freedom in New General Relativity. 
This theory is the three-parameter extension of Teleparallel Equivalent to GR and classified into nine irreducible types according to the rotation symmetry $SO(3)$ on each leaf of ADM-foliation. 
In the previous work~[{\it Phys. Rev. D 112 (2025) 8, 084052}], we investigated the degrees of freedom in NGR types that are of interest in describing gravity: Type 2, Type 3, Type 5, and Type 8. 
In this work, we focus on unveiling those numbers in all other types to complete the analysis of NGR. 
After providing the Hamiltonian formulation of NGR and considering in detail the regularity of NGR, we perform the analysis of Type 4, Type 7, and Type 9. 
We reveal that the degrees of freedom of Type 4, Type 7, and Type 9 are five, zero (purely topological system in bulk spacetime), and three, respectively. 
Type 4 and Type 9 have second-class constraint densities only, whereas Type 7 has first-class constraint densities only. 
In every type, no bifurcation occurs. 
In particular, Type 4 and Type 7 are irregular and provide specific examples of handling irregular systems. 
Since no general method is known for treating an irregular system, this work contributes to furthering the understanding of irregular systems. 
\end{abstract}

\maketitle

\section{\label{01}Introduction}
Metric-Affine gauge theories of Gravity (MAG) are a candidate for going beyond General Relativity (GR) to reconcile the recent observational issues in cosmology such as inflation~\cite{Planck:2018vyg,Vazquez:2018qdg,Tsujikawa:2003jp}, dark matter~\cite{Freese:2008cz,Planck:2018vyg,Billard:2021uyg}, dark energy (the late-time accelerated expansion of the universe)~\cite{SupernovaSearchTeam:1998fmf,SupernovaCosmologyProject:1998vns,Planck:2018vyg}, the tension problem in cosmological parameters~\cite{Planck:2018vyg,H0LiCOW:2019pvv,Riess:2019cxk,Schoneberg:2022ggi,ACT:2023kun}. 
(See also Sec.I of Ref.~\cite{Tomonari:2024ybs} for details.) 
MAG is constructed upon two principles, {\it i.e.,} the principle of frame and diffeomorphism symmetry, utilizing the metric-affine geometry in terms of bundle theories, in which a spacetime is described by a differential manifold equipped with a metric tensor and an affine connection independently~\cite{Utiyama:1956sy,Kibble:1961ba,Ivanenko:1983fts,Hehl:1994ue,Tomonari:2023ars}. 
MAG is composed of independent classes according to gauge conditions~\cite{Hehl:1994ue,Blagojevic:2002du,Adak:2005cd,Adak:2006rx,Adak:2008gd,Adak:2011ltj,Bahamonde:2021gfp,Tomonari:2023ars}. 
In a set of gauge conditions for vanishing non-metricity, MAG leads to GR and TEGR, in which gravity is described by using the curvature of the Levi-Civita connection and the torsion under the imposition of vanishing generic curvature (teleparallel condition), respectively~\cite{Bahamonde:2021gfp}.
A hybrid theory of these geometric quantities is known as Poincare's gauge theory of gravity~\cite{Hehl:1976kj,Hayashi:1981fx,Blagojevic:2003cg,Hehl:2023khc}. 

TEGR can be extended using three-free parameters into New General Relativity (NGR)~\cite{Hayashi:1979qx}. 
(See also II-A of Ref.~\cite{Tomonari:2024ybs} for details.) 
NGR is classified into nine independent types according to the $SO(3)$-irreducible decomposition of canonical momenta in four-dimensional spacetime~\cite{Blixt:2018znp,Pati:2022nwi}. 
Counting Degrees of Freedom (DoF) of NGR is the purpose of this series of papers. 
This is an important subject for the applications of perturbative theories to cosmology and astrophysics, since only when the strong couplings, which only occurs in the perturbative theory of a gravity theory and never indicates the fault of the gravity theory itself, and ghost modes are absent, the perturbative theory predicts and explains physical phenomena in a healthy manner~\cite{Gomes:2023tur,Bahamonde:2024zkb}. 
We note here that strong coupling can be remedied by hand, whereas ghost mode is a fatal issue. 
As examples for the former statement, there are lessons of massive gravity and massive Yang-Mills theory~\cite{Vainshtein:1972sx,Hell:2021oea}.  

Unveiling the DoFs of a given theory relate deeply to the investigation of these issues. 
In the previous work~\cite{Tomonari:2024ybs}, we counted the non-linear DoFs of NGR focusing on Type 2, Type 3, Type 5, and Type 8, applying the Dirac-Bergmann analysis~\cite{Dirac:1950pj,Dirac:1958sq,Bergmann:1949zz,BergmannBrunings1949,Bergmann1950,Anderson:1951ta}. 
These types contain gravitational propagating modes (= tensor modes)~\cite{Bahamonde:2024zkb}. 
In particular, in the previous work~\cite{Tomonari:2024ybs}, we also unveiled that a set of specific conditions on Lagrange multipliers cause a bifurcation, and we define a new class of constraint densities labeled as a semi-first class constraint density. 
In Type 8, the semi-first class constraint densities exist and bifurcations occur.
Here, bifurcation means a branching in the emergence of constraints, and it occurs in such a case that a coefficient in the Poisson bracket algebra (PB-algebra) vanishes under some condition. 

In this paper, we focus on the remaining types, {\it i.e.,} Type 4, Type 7, and Type 9. 
These types do not contain gravitational propagating modes~\cite{Bahamonde:2024zkb}; 
these types are ruled out as theories of gravity in applications to cosmology.
However, as will be briefly mentioned in conclusion section, these types would still provide dynamical models in other research fields. 
Furthermore, we show that Type 4 and Type 7 are irregular systems, --- a system such that the functional independence of constraints is violated ---, and describe specific treatment of these systems.
Since no general method exists to treat irregular systems, it is essential to collect as many examples as possible, with the goal of developing a comprehensive approach toward manipulating irregular systems.
We will also show that in these types, all constraint densities are classified either as second-class (Type 4 and Type 9) or first-class (Type 7). 
Bifurcations do not occur in any of these types;
there is no room for any semi-first class constraint densities to appear, unlike in Type 8~\cite{Tomonari:2024ybs}. 

The construction of this paper is given as follows. 
In Sec.~\ref{02}, we summarize the fundamental ingredients of NGR in the $SO(3)$-irreducible decomposition of canonical momentum to perform the analysis. 
In Sec.~\ref{03}, based on the method provided in the previous work~\cite{Tomonari:2024ybs}, the Dirac-Bergmann analysis is performed on each type of NGR. 
First, we consider the regularity of each type of NGR and clarify that only Type 7 and Type 9 are irregular. 
Second, we reveal that the DoF of Type 4, Type 7, and Type 9 are five, zero (purely topological system in bulk spacetime), and three, respectively. 
No bifurcation occurs, unlike to the case of Type 8 in the previous work~\cite{Tomonari:2024ybs}. 
Finally, in Sec.~\ref{04}, we conclude this series of works. 
Throughout this paper, we use units and notation according to the previous work~\cite{Tomonari:2024ybs}.

\section{\label{02}NGR Hamiltonian and primary constraints}
We briefly summarize NGR in the Hamiltonian formulation. 
The reformulation of NGR based on the gauge approach to gravity and the ADM-foliation of NGR are provided in detail in our previous work: Ref.~\cite{Tomonari:2024ybs}. 
We assume that the dimension of spacetime is four. 

The configuration space $\mathcal{Q}$ is spanned by lapse function $\alpha$, shift vector $\beta^{i}$, and co-frame field components $\theta^{A}{}_{i}$. 
As discussed in Sec.~III of Ref.~\cite{Blixt:2018znp}, as long as we consider the theory in Weitzenb\"{o}ck gauge, the Hamiltonian is independent from $\Lambda^{A}{}_{B}$ and $\hat{\pi}^{AB}\,$. 
Then canonical momentum with respect to $\theta^{A}{}_{i}$ is derived as follows~\cite{Blixt:2018znp}:
\begin{equation}
    \pi_{A}{}^{i}={}^{\mathcal{V}}\pi^{i}\xi_{A}+{}^{\mathcal{A}}\pi^{ji}h_{kj}\theta_{A}{}^{k}+{}^{\mathcal{S}}\pi^{ji}h_{kj}\theta_{A}{}^{k}+{}^{\mathcal{T}}\pi\theta_{A}{}^{i}\,,
\label{SO(3)-irreducible decomposition of canonical momenta}
\end{equation}
where ${}^{\mathcal{V}}\pi^{i}$, ${}^{\mathcal{A}}\pi^{ji}$, ${}^{\mathcal{S}}\pi^{ji}$, and ${}^{\mathcal{T}}\pi$ are the vectorial, anti-symmetric, symmetric trace-free and trace part of the momentum $\pi_{A}{}^{i}$, which are given as follows~\cite{Pati:2022nwi}:
\begin{equation}
\begin{split}
    &{}^{\mathcal{V}}\pi^{i}=-\xi^{A}\pi_{A}{}^{i}\,,\quad{}^{\mathcal{A}}\pi^{ij}=\pi^{[ij]}=-\frac{1}{2}\pi_{A}{}^{i}\theta^{A}{}_{k}h^{jk}+\frac{1}{2}\pi_{A}{}^{j}\theta^{A}{}_{k}h^{ik}\,,\\
    &{}^{\mathcal{S}}\pi^{ij}=\pi^{(ij)}-\frac{1}{3}\pi_{A}{}^{k}\theta^{A}{}_{k}h^{ij}=\frac{1}{2}\pi_{A}{}^{i}\theta^{A}{}_{k}h^{jk}+\frac{1}{2}\pi_{A}{}^{j}\theta^{A}{}_{k}h^{ik}-\frac{1}{3}\pi_{A}{}^{k}\theta^{A}{}_{k}h^{ij}\,,\quad{}^{\mathcal{T}}\pi=\frac{1}{3}\pi_{A}{}^{i}\theta^{A}{}_{i}\,.
\end{split}    
\label{SO3-irreducible canonical momenta}
\end{equation}
Canonical momenta with respect to the lapse function and the shift vector are $\pi_{0} = 0$ and $\pi_{i} = 0$. 

The total-Hamiltonian of NGR is provided as follows~\cite{Blixt:2018znp}:
\begin{equation}
\begin{split}
    \mathcal{H} = &\Big({}^\mathcal{V}\mathcal{H} + {}^\mathcal{A}\mathcal{H} + {}^\mathcal{S}\mathcal{H} + {}^\mathcal{T}\mathcal{H} \Big) + D_i[\pi_A{}^i(\alpha \xi^{A}+\beta^{j}\theta^A{}_j )]\\
    &\quad\quad\quad\quad - \alpha \Big(\sqrt{h}\ {}^{3}\mathbb{T} - \xi^A D_i\pi_A{}^i \Big) - \beta^k\Big(T^A{}_{jk}\pi_A{}^j + \theta^A{}_k D_{i}\pi_A{}^i\Big) + {}^{\alpha}\lambda\pi_{0} + {}^{\beta}\lambda^{i}\pi_{i}\,,
\end{split}
\label{}
\end{equation}
where 
\begin{align}
&{}^\mathcal{V} \mathcal{H} =
\begin{cases}
\alpha \sqrt{h}\ \frac{{}^{\mathcal{V}}C_i {}^{\mathcal{V}}C^i}{4 A_{\mathcal{V}}}& \textrm{ for } {}^{\mathcal{V}}A \neq 0\\
\sqrt{h}\ {}^\mathcal{V}\lambda_i {}^{\mathcal{V}}C^i & \textrm{ for } {}^{\mathcal{V}}A = 0\,,
\end{cases}
\quad
{}^\mathcal{A} \mathcal{H} =
\begin{cases}\label{VH and AH}
-\alpha \sqrt{h}\ \frac{{}^{\mathcal{A}}C_{ij} {}^{\mathcal{A}}C^{ij}}{4 A_{\mathcal{A}}}& \textrm{ for } {}^{\mathcal{A}}A \neq 0\\
\sqrt{h}\ {}^\mathcal{A}\lambda_{ij} {}^{\mathcal{A}}C^{ij} & \textrm{ for } {}^{\mathcal{A}}A = 0\,,
\end{cases}\\
&{}^\mathcal{S} \mathcal{H} =
\begin{cases}
-\alpha \sqrt{h}\ \frac{{}^{\mathcal{S}}C_{ij} {}^{\mathcal{S}}C^{ij}}{4 A_{\mathcal{S}}}& \textrm{ for } {}^{\mathcal{S}}A \neq 0\\
\sqrt{h}\ {}^\mathcal{S}\lambda_{ij} {}^{\mathcal{S}}C^{ij} & \textrm{ for } {}^{\mathcal{S}}A = 0\,,
\end{cases}
\quad
{}^\mathcal{T} \mathcal{H} =
\begin{cases}\label{SH and TH}
-\alpha \sqrt{h}\ \frac{3{}^{\mathcal{T}}C {}^{\mathcal{T}}C}{4 A_{\mathcal{T}}}& \textrm{ for } {}^{\mathcal{T}}A \neq 0\\
\sqrt{h}\ {}^\mathcal{T}\lambda {}^{\mathcal{T}}C & \textrm{ for } {}^{\mathcal{T}}A = 0\,,
\end{cases}
\end{align}
where ${}^{\mathcal{A}}\lambda_{ij}$ and ${}^{\mathcal{S}}\lambda_{ij}$ are not symmetric and not anti-symmetric with respect to the indices $i$ and $j$, respectively, and
\begin{equation}
    {}^{3}\mathbb{T}=c_{1}\eta_{AB}T^A{}_{ij}T^B{}_{kl}h^{ik}h^{jl}+c_{2}\theta_A{}^i \theta_B{}^j T^A{}_{kj}T^B{}_{li}h^{kl}+c_{3}\theta_A{}^i\theta_B{}^j h^{kl}T^A{}_{ki}T^B{}_{lj}\,.
\label{}
\end{equation}
The cases for ${}^{\mathcal{V}}A=0$, ${}^{\mathcal{A}}A=0$, ${}^{\mathcal{S}}A=0$, and/or ${}^{\mathcal{T}}A=0$ are nothing but constraints, and then the variables ${}^\mathcal{V}\mathcal{C}^i$, ${}^\mathcal{A}\mathcal{C}^{ij}$, ${}^\mathcal{S}\mathcal{C}^{ij}$, and ${}^\mathcal{T}\mathcal{C}$ turn to be primary constraints given as follows~\cite{Blixt:2018znp}:
\begin{equation}
\begin{split}
    &{}^\mathcal{V}\mathcal{C}^i := \frac{{}^{\mathcal{V}}\pi^{i}}{\sqrt{h}} - 2  c_{3}T^B{}_{kl} h^{ik}\theta_B{}^l \approx 0\,\quad (A_{\mathcal{V}} = 0)\,,\quad{}^\mathcal{A}\mathcal{C}^{ij} := \frac{{}^\mathcal{A}\pi^{ij}}{\sqrt{h}} - 2  c_{2} h^{li}h^{jk}T^B{}_{kl} \xi_{B} \approx 0\,\quad (A_{\mathcal{A}} = 0)\,,\\
    &{}^\mathcal{S}\mathcal{C}^{ij} :=  \frac{{}^\mathcal{S}\pi^{ij}}{\sqrt{h}} \approx 0\,\quad (A_{\mathcal{S}} = 0)\,,\quad{}^\mathcal{T}\mathcal{C} :=  \frac{{}^\mathcal{T}\pi}{\sqrt{h}} \approx 0\,\quad (A_{\mathcal{T}} = 0)\,.    
\end{split}
\label{SO3-irreducible primary constraints}
\end{equation}
The non-constraint cases that are given in the non-vanishing ${}^{\mathcal{V}}A$, ${}^{\mathcal{A}}A$, ${}^{\mathcal{S}}A$, and/or ${}^{\mathcal{T}}A$, the explicit forms of these variables are provided in Ref.~\cite{Blixt:2018znp}. 
According to the $SO(3)$-irreducible decomposition, we can classify the theory into nine types as shown in Table~\ref{Types of NGR}~\cite{Blixt:2018znp}. 
\begin{table}[ht!]
    \centering
    \renewcommand{\arraystretch}{1.5}
    \begin{tabular}{ c || c | c }
        Theory & Conditions & $SO(3)$-irreducible primary constraints \\ \hline\hline
        Type 1 & $A_{I}\neq 0 \ \forall I\in \{ \mathcal{V},{\mathcal{A}},{\mathcal{S}},\mathcal{T} \}$ &  No constraint\\ \hline
        Type 2 & $A_{\mathcal{V}}=0$ &  ${}^{\mathcal{V}}\mathcal{C}_{i}\approx0$\\ \hline
        Type 3 & $A_{\mathcal{A}}=0$ &   ${}^{\mathcal{A}}\mathcal{C}_{ij}\approx0$\\ \hline
        Type 4 & $A_{\mathcal{S}}=0$ &   ${}^{\mathcal{S}}\mathcal{C}_{ij}\approx0$\\ \hline
        Type 5 & $A_{\mathcal{T}}=0$ &   ${}^{\mathcal{T}}\mathcal{C}\approx0$\\ \hline
        Type 6 & $A_{\mathcal{V}}=A_{\mathcal{A}}=0$ & ${}^{\mathcal{V}}\mathcal{C}_{i}={}^{\mathcal{A}}\mathcal{C}_{ij}\approx0$\\ \hline
        Type 7 & $A_{\mathcal{A}}=A_{\mathcal{S}}=0$ &  ${}^{\mathcal{A}}\mathcal{C}_{ij}={}^{\mathcal{S}}\mathcal{C}_{ij}\approx0$\\ \hline
        Type 8 & $A_{\mathcal{A}}=A_{\mathcal{T}}=0$ &  ${}^{\mathcal{A}}\mathcal{C}_{ij}={}^{\mathcal{T}}C\approx0$\\ \hline
        Type 9 & $A_{\mathcal{V}}=A_{\mathcal{S}}=A_{\mathcal{T}}=0$ &  ${}^{\mathcal{V}}\mathcal{C}_{i}={}^{\mathcal{S}}\mathcal{C}_{ij}={}^{\mathcal{T}}C\approx0$\\ \hline
    \end{tabular}
    \caption{All types of NGR in the $SO(3)$-irreducible decomposition of canonical momentum. Type 6 is TEGR.}
    \label{Types of NGR}
\end{table}

\section{\label{03}Hamiltonian analysis of Type 4, Type 7, and Type 9}
\subsection{\label{03:01}Common sector: Diffeomorphism symmetry}
As investigated in Ref.~\cite{Tomonari:2024ybs}, the diffeomorphism sector has the common PB-algebra without depending on the content of each constraint in Type 2, Type 3, Type 5, and Type 8 of NGR. 
The same statement holds in all the remaining types. That is, the total-Hamiltonian to each type is given as follows:
\begin{equation}
\begin{split}
    \mathcal{H}_{\rm Type\,4} = & \sqrt{h}{}^{\mathcal{S}}\lambda_{ij}(T^{*}\mathcal{Q}){}^{\mathcal{S}}\mathcal{C}^{ij} + D_i[\pi_A{}^i(\alpha \xi^{A}+\beta^{j}\theta^A{}_j )]\\
    & + \alpha \Big( - \sqrt{h}\ {}^{3}\mathbb{T} - \xi^A D_i\pi_A{}^i + \sqrt{h}\ \frac{{}^{\mathcal{V}}\mathcal{C}_i {}^{\mathcal{V}}\mathcal{C}^i}{4 A_{\mathcal{V}}} - \sqrt{h}\ \frac{{}^{\mathcal{A}}\mathcal{C}_{ij} {}^{\mathcal{A}}\mathcal{C}^{ij}}{4 A_{\mathcal{A}}} - \sqrt{h}\ \frac{3{}^{\mathcal{T}}\mathcal{C} {}^{\mathcal{T}}\mathcal{C}}{4 A_{\mathcal{T}}}\Big) + {}^{\alpha}\lambda\pi_{0}\\
    & + \beta^k\Big( - T^A{}_{jk}\pi_A{}^j - \theta^A{}_k D_{i}\pi_A{}^i\Big) + {}^{\beta}\lambda^{i}\pi_{i}\,,
\end{split}
\label{total Hamiltonian of type 4}
\end{equation}

\begin{equation}
\begin{split}
    \mathcal{H}_{\rm Type\,7} =& \sqrt{h}{}^{\mathcal{A}}\lambda_{ij}(T^{*}\mathcal{Q}){}^{\mathcal{A}}\mathcal{C}^{ij} + \sqrt{h}{}^{\mathcal{S}}\lambda_{ij}(T^{*}\mathcal{Q}){}^{\mathcal{S}}\mathcal{C}^{ij} + D_i[\pi_A{}^i(\alpha \xi^{A}+\beta^{j}\theta^A{}_j )]\\
    & + \alpha \Big( - \sqrt{h}\ {}^{3}\mathbb{T} - \xi^A D_i\pi_A{}^i + \sqrt{h}\ \frac{{}^{\mathcal{V}}\mathcal{C}_i {}^{\mathcal{V}}\mathcal{C}^i}{4 A_{\mathcal{V}}} - \sqrt{h}\ \frac{3{}^{\mathcal{T}}\mathcal{C} {}^{\mathcal{T}}\mathcal{C}}{4 A_{\mathcal{T}}}\Big) + {}^{\alpha}\lambda\pi_{0}\\
    & + \beta^k\Big( - T^A{}_{jk}\pi_A{}^j - \theta^A{}_k D_{i}\pi_A{}^i\Big) + {}^{\beta}\lambda^{i}\pi_{i}\,,
\end{split}
\label{total Hamiltonian of type 7}
\end{equation}

\begin{equation}
\begin{split}
    \mathcal{H}_{\rm Type\,9} =& \sqrt{h}{}^{\mathcal{V}}\lambda_{i}(T^{*}\mathcal{Q}){}^{\mathcal{V}}\mathcal{C}^{i} + \sqrt{h}{}^{\mathcal{S}}\lambda_{ij}(T^{*}\mathcal{Q}){}^{\mathcal{S}}\mathcal{C}^{ij} + \sqrt{h}{}^{\mathcal{T}}\lambda(T^{*}\mathcal{Q}){}^{\mathcal{T}}\mathcal{C} + D_i[\pi_A{}^i(\alpha \xi^{A}+\beta^{j}\theta^A{}_j )]\\
    & + \alpha \Big( - \sqrt{h}\ {}^{3}\mathbb{T} - \xi^A D_i\pi_A{}^i - \sqrt{h}\ \frac{{}^{\mathcal{A}}\mathcal{C}_{ij} {}^{\mathcal{A}}\mathcal{C}^{ij}}{4 A_{\mathcal{S}}}\Big) + {}^{\alpha}\lambda\pi_{0}\\
    & + \beta^k\Big( - T^A{}_{jk}\pi_A{}^j - \theta^A{}_k D_{i}\pi_A{}^i\Big) + {}^{\beta}\lambda^{i}\pi_{i}\,.
\end{split}
\label{total Hamiltonian of type 9}
\end{equation}
The coefficients of the lapse function, $\alpha$, and the shift vector, $\beta^{i}$, are primary constraints. 
These primary constraints form the hypersurface deformation algebra~\cite{Dirac:1958sc} regardless of the type of NGR. 
In addition, we can show that these primary constraints commute with the remaining primary constraints. 
Therefore, we investigate the constraint structures formed by the first line of each total Hamiltonian~\cite{Tomonari:2024ybs}. 

\subsection{\label{03:01Added}Regularity of NGR}
To apply the Dirac-Bergmann analysis in a straightforward way, the theory must satisfy the regularity condition, \textit{i.e.}, the first-order variation of every constraint is expressed as a linear combination of constraints existing in the given theory, as stated in Ref.~\cite{Dirac:1950pj,Dirac:1958sc,Dirac:1958sq}. 
This condition is also equivalent to the functional independence of constraints around the constraint surface.\footnote{
Assume that constraints $C_{a}\approx0$ $(a = 1,2,3,\cdots,r)$ exist. 
The independence of $C_{a}$ around the constraint surface $\Gamma=\{(q_{i},p^{i}) |C_{a}(q_{i},p^{i}) = 0\} \subset S$, where $S$ is a symplectic manifold with dimension $2n$ larger than $r$, is defined by the satisfaction of the following two conditions:
\begin{equation}
    \left.\sum^{r}_{a=1} f_{a} \delta C_{a} \right|_{\Gamma} = 0, 
\end{equation}
for arbitrary $f_{a}$, and
\begin{equation}
    \left.\sum^{r}_{a=1} f_{a} \delta C_{a}\right|_{S \cap {}^\lnot\Gamma} = 0
\end{equation}
around $\Gamma$, not on $\Gamma$ itself, only for $f_{a} = 0$. 
Here, we denote $\delta C_{a}$ as the first-order variation of $C_{a}$. 
If $\delta C_{a}$ can be expressed in terms of the linear combination of the constraints $C_{a}$, the definition holds. 
The converse is also true. 
}
In Ref.~\cite{Miskovic:2003ex}, an irregular constraint is classified as Type I (Multi-linear constraints) and Type II (Non-linear constraints).
That is, we call a constraint
\begin{equation}
    \phi\equiv\prod_{i=1}^{M}f_{i}(z) \approx0\,
\label{classification of irregular constraints}
\end{equation}
multi-linear constraint, where $f_{i} = f_{i}(z)$ are the functions of the $2n$-dimensional phase-space variables $z = (z^{1}\,,z^{2}\,,\cdots,z^{2n})\,$ such that $f_{i}$ vanishes in some region of the constraint surface. 
If all $f_{i}$ are identical, {\it i.e.,} $f_{i} = f$ for all $i = 1\,,2\,,\cdots\,,M(>1)\,$, we call Eq.~(\ref{classification of irregular constraints}) non-linear constraint. 
In this case, $\phi$ becomes a first-class constraint.\footnote{
For another constraint $\psi \approx 0\,$, 
\begin{equation}
    \{\phi\,,\psi\} = \{f^{M}\,,\psi\} = f \{f^{M-1}\,,\psi\} + \{f\,,\psi\} f^{M-1} \approx0\,;
\end{equation}
on a region where $f$ vanishes. Thus, a constraint classified as Type II $M > 1$ is always first-class constraint.
}

A regularity condition is classified into three cases; 
A, The condition holds on the entire constraint surface; 
B, The condition does not hold on the entire constraint surface; 
C, The condition holds on a part of the constraint surface. 
For each type of irregular constraints, in Ref.~\cite{Miskovic:2003ex}, the authors provide two examples of point particle systems with $M = 2$ and a consideration of Chern-Simons theory. 
In particular, in the latter example, the author unveils that when an irregular system is linearized, the resulting system can raise an extra DoF, and this DoF is not ascribed to the original theory; 
it is just an accidental emergence due to the irregularity of the original system. 
In other words, for an irregular system, we cannot compare DoFs that are identified by the Dirac-Bergmann analysis on a given system with those in the linearized system. 
(In detail, see Ref.~\cite{Miskovic:2003ex}.)
However, there is no general procedure to manipulate irregular systems; 
we must manipulate each system by each way. 

To count out the number of DoFs in irregular systems using the Dirac-Bergmann analysis, we have to regularize the system. 
That is, we reformulate a set of irregular constraints such that a new set of constraints is regular.
However, this treatment may change the dynamics of the original system; 
there is no general procedure to guarantee the equivalence of the dynamics when we regularize an irregular system. 
Fortunately, if a regularization process does not change the PB-algebra, the dynamics always holds even after regularizing the constraints at least on a partial region of the original constraint surface. 
We denote the original constraint surface and the regularized one by $\Gamma_{0}$ and $\Gamma_{1}$. 
If $\Gamma_{01}:=\Gamma_{0}\cap\Gamma_{1}$ is not empty, on $\Gamma_{01}$, the dynamics of the regularized system is equivalent to that of the original one. 
If $\Gamma_{01}$ is empty, the regularized system is completely a different constraint system. 
Since a PB-algebra characterizes a constraint system, if the algebra holds, $\Gamma_{01}$ is not empty. 
Although there is no general procedure, we use this fact to compare the regularized system to the original one from the point of view of the coincidence of dynamics. 
In this work, we perform the Dirac-Bergmann analysis based on this perspective.

In NGR theory, as stated in Sec.~III-A of Ref.~\cite{Tomonari:2024ybs}, there is a doubt that the regularity condition of the diagonal components in ${}^\mathcal{S}\mathcal{C}^{ij} \propto {}^{\mathcal{S}}\pi^{ij}$ holds. 
That is, if the trace-free condition of ${}^\mathcal{S}\mathcal{C}^{ij}$ does not satisfy as strong equality then the regularity is violated. 
This situation could be anticipated by the fact that the $SO(3)$-irreducible decomposition of canonical momenta, Eq.~(\ref{SO(3)-irreducible decomposition of canonical momenta}), holds in a well-defined manner only when the trace of ${}^{\mathcal{S}}\pi^{ij}$ vanishes as a strong equality. 
Otherwise, the functional degrees of freedom in the left-hand side in Eq.~(\ref{SO(3)-irreducible decomposition of canonical momenta}), {\it i.e.}, the number of twelve, does not match that number on the right side, {\it i.e.}, the number of thirteen. 
To check this more precisely, according to Dirac's work~\cite{Dirac:1950pj}, let us calculate the first-order variation of the trace of ${}^\mathcal{S}\mathcal{C}^{ij}$ as follows:
\begin{equation}
    \delta(h_{ij}{}^\mathcal{S}\mathcal{C}^{ij}) = \left(-h_{ik}h_{jl}\delta h^{kl} + \frac{1}{2}h_{kl}h_{ij}\delta h^{kl}\right)\cdot{}^\mathcal{S}\mathcal{C}^{ij} - h_{kl}\delta h^{kl}\cdot{}^\mathcal{T}\mathcal{C}
\label{1st-order variation of Tr SC}
\end{equation}
Here, notice that the variation of the metric on a leaf $\Sigma_{t}$ can be decomposed into those of the co-frame field components $\theta^{A}{}_{i}\,$; 
the above equations consist only of the phase-space variable as desired. 
Then, Eq.~(\ref{1st-order variation of Tr SC}) indicates that the trace-free property holds only when both ${}^\mathcal{S}\mathcal{C}^{ij}$ and ${}^{\mathcal{T}}C$ become constraints.
In other words, if $\delta(h_{ij}{}^\mathcal{S}\mathcal{C}^{ij})$ depends only on ${}^\mathcal{S}\mathcal{C}^{ij}$, all types of NGR always satisfy the regularity condition. 
This consideration leads to the following property:\\ \\ 
\hypertarget{lem01}{\textbf{Regularity Condition (RC):}}\\
{\it{
The trace of ${}^\mathcal{S}\mathcal{C}^{ij}$ is regular if and only if both ${}^\mathcal{S}\mathcal{C}^{ij}$ and ${}^\mathcal{T}\mathcal{C}$ form the constraints of the theory.
}}
\begin{flushright}$\blacksquare$\end{flushright}
This property states that ${}^\mathcal{T}\mathcal{C}$ should be a constraint in order to make the diagonal components of ${}^\mathcal{S}\mathcal{C}^{ij}$ regular. 
We note that we use this fact to regularize Type 4 of NGR.
\hyperlink{lem01}{RC} and the considerations in Ref.~\cite{Tomonari:2024ybs} (Type 2, Type 3, Type 5, Type 6, Type 8 do not contain the constraint ${}^\mathcal{S}\mathcal{C}^{ij} \approx 0$) lead us to the following theorem:\\ \\
\hypertarget{thm01}{\textbf{Regularity of NGR:}}\\
{\it{
Each Type of NGR is classified as follows:\\
(a). Type 2, Type 3, Type 5, Type 6, Type 8, and Type 9 are regular: Case A in Type I with $M=1$.\\
(b). Type 4 is irregular: Case C in Type I with $M=1$.\\
(c). Type 7 is irregular: Case B in Type I with $M=1$.
}}
\begin{flushright}$\blacksquare$\end{flushright}
We will show the attribute of Case A, B, or C associated with Type 4 and Type 7 in the following sections.
We note that Type 1 is out of consideration since Type 1 does not contain any constraints. 
Therefore, in the current work, the Dirac-Bergmann analysis can be straightforwardly applied only to Type 9. 
We have to regularize Type 4 and Type 7 to count out the number of DoFs.
As mentioned in Ref.~\cite{Tomonari:2024ybs}, in these irregular systems, the trace-free condition of ${}^\mathcal{S}\mathcal{C}^{ij}$ holds only as a weak equality, indicating that only two of the three diagonal components of ${}^\mathcal{S}\mathcal{C}^{ij}$ are linearly independent. 

Now we are ready to analyze Type 4, Type 7, and Type 9.

\subsection{\label{03:02}Specific sector in Type 4: DoF in Type 4}
The part of governing the proper symmetry in Type 4 of NGR is given by
\begin{equation}
     \tilde{\mathcal{H}}_{\rm Type\,4} = \sqrt{h}\,{}^{\mathcal{S}}\lambda_{ij}(T^{*}\mathcal{Q})\,{}^{\mathcal{S}}\mathcal{C}^{ij} + D_i[\pi_A{}^i(\alpha \xi^{A}+\beta^{j}\theta^A{}_j )]\,.
\label{specific part of H_type4}
\end{equation}
In this type of NGR, the primary constraints are ${}^{\mathcal{S}}\mathcal{C}^{ij} \approx 0$. 
As mentioned in Sec.~III B of Ref.~\cite{Tomonari:2024ybs} and Ref.~\cite{Tomonari:2023wcs}, the total divergent term in Eq.~(\ref{specific part of H_type4}) can be discarded by virtue of the diffeomorphism symmetry. 
Therefore, the DoF of this type is determined only by the constraint structure of ${}^{\mathcal{S}}\mathcal{C}^{ij} \approx 0$. 
To investigate this, we first need the PB-algebra of the primary constraints as follows:
\begin{equation}
\begin{split}
    &\{{}^{\mathcal{S}}C^{ij}(t\,,\vec{x})\,,{}^{\mathcal{S}}C^{kl}(t\,,\vec{y})\} = \\
    &\quad\quad\quad\quad \frac{1}{\sqrt{h}}\,\left[h^{ik}\,{}^{\mathcal{A}}C^{jl} + h^{il}\,{}^{\mathcal{A}}C^{jk} + h^{jk}\,{}^{\mathcal{A}}C^{il} + h^{jl}\,{}^{\mathcal{A}}C^{ik}\right]\,\delta^{(3)}(\vec{x} - \vec{y}) - \frac{2c_{2}}{\sqrt{h}}\,\xi_{A}\,T^{A}{}_{ab}\,{}^{\mathcal{S}}H^{ijklab}\,\delta^{(3)}(\vec{x} - \vec{y})\,,\\
\end{split}
\label{PB-algebra of SC and SC with upper indices}
\end{equation}
where we set
\begin{equation}
    {}^{\mathcal{S}}H^{ijklab} := 2\,h^{ja}\,h^{i(k}\,h^{l)b} + 2\,h^{ia}\,h^{j(k}\,h^{l)b}\,.
\label{VH^ijklab}
\end{equation}
Apparently, this PB-algebra indicates that the primary constraints ${}^{\mathcal{S}}\mathcal{C}^{ij} \approx 0$ are classified into second-class constraints.
Notice that the last term in Eq.~(\ref{PB-algebra of SC and SC with upper indices}) does not vanish even if integrating over with respect to both space coordinates $\vec{x}$ and $\vec{y}$, which is different from the case in Type 8~\cite{Tomonari:2024ybs}. 
Therefore, the primary constraints ${}^{\mathcal{S}}\mathcal{C}^{ij} \approx 0$ are classified into second-class constraints in terms both of density and smeared variables. 
As mentioned in Sec.~\ref{03:01Added}, the irregular property of Type 4 indicates that two out of three diagonal components of ${}^\mathcal{S}\mathcal{C}^{ij} \propto {}^{\mathcal{S}}\pi^{ij}$ are regular constraints. 
(See \hyperlink{thm01}{Regularity of NGR - (b).}) 
Thus, only five out of six multipliers are determined by the consistency conditions, and the remaining one is still arbitrary. 
However, Eq.~(\ref{1st-order variation of Tr SC}) indicates the existence of a second-class constraint that is related to the diagonal components of ${}^{\mathcal{S}}\mathcal{C}^{ij}$. 
As mentioned in Sec.~\ref{03:01Added}, \hyperlink{lem01}{RC} implies that if it is possible to switch ${}^{\mathcal{T}}C$ to a constraint then Type 4 turns into a regular system.
Taking into account the following PB-algebra,
\begin{equation}
    \{{}^{\mathcal{S}}C^{ij}(t\,,\vec{x})\,,{}^{\mathcal{T}}C(t\,,\vec{y})\} = - \frac{4}{3}\,\frac{1}{\sqrt{h}}\,{}^{\mathcal{S}}C^{ij}\,\delta^{(3)}(\vec{x} - \vec{y}) - \frac{1}{3}\,\frac{1}{h}\,\theta^{A}{}_{k}\,\pi_{A}{}^{(i}\,h^{j)k}\,\delta^{(3)}(\vec{x} - \vec{y})\,,
\label{PB-algebra of SC and TC with upper indices}
\end{equation}
${}^{\mathcal{T}}C$ should be a {\it secondary} second-class constraint, and this is nothing but the remaining one we are looking for determining the remaining multiplier. 
This statement is consistent with the fact that the total number of second-class constraints is always even~\cite{Tomonari:2023vgg}. 
Summarizing, we count out the DoF of Type 4 as follows: ${\rm DoF} = (16\times2 - 8\times2 - 5 - 1)/2 = 5\,$.

This result is generically valid only on a subsurface of the constraint surface of the origin system due to the imposition of ${}^{\mathcal{T}}C \approx0\,$. 
That is, the result is valid only on $\Gamma_{1}\cap\Gamma_{0}$, which is not empty since $\Gamma_{1}\subset\Gamma_{0}$ holds; Type 4 is classified as Case C in Type I.
For the other region of the origin constraint surface, we need to consider another regularization to analyze this system and count the degrees of freedom, although there is no generic procedure.

\subsection{\label{03:03}Specific sector in Type 7: DoF in Type 7}
The part of governing the proper symmetry in Type 7 of NGR is given by
\begin{equation}
     \tilde{\mathcal{H}}_{\rm Type\,7} = \sqrt{h}\,{}^{\mathcal{A}}\lambda_{ij}(T^{*}\mathcal{Q})\,{}^{\mathcal{A}}\mathcal{C}^{ij} + \sqrt{h}\,{}^{\mathcal{S}}\lambda_{ij}(T^{*}\mathcal{Q})\,{}^{\mathcal{S}}\mathcal{C}^{ij} + D_i[\pi_A{}^i(\alpha \xi^{A}+\beta^{j}\theta^A{}_j )]\,.
\label{specific part of H_type7}
\end{equation}
The DoF of this type of NGR is determined by the constraint structure of ${}^{\mathcal{A}}\mathcal{C}^{ij} \approx 0$ and ${}^{\mathcal{S}}\mathcal{C}^{ij} \approx 0$.
The PB-algebra of these constraints are calculated as follows:
\begin{equation}
\begin{split}
    &\{{}^{\mathcal{S}}C^{ij}(t\,,\vec{x})\,,{}^{\mathcal{A}}C^{kl}(t\,,\vec{y})\} = \left[\frac{1}{2\sqrt{h}}\,\left(h^{ik}\,{}^{\mathcal{A}}C^{jl} - h^{il}\,{}^{\mathcal{A}}C^{jk} + h^{jk}\,{}^{\mathcal{A}}C^{il} - h^{jl}\,{}^{\mathcal{A}}C^{ik} + \frac{4}{3}\,h^{ij}\,{}^{\mathcal{A}}C^{kl}\right)\right.\\
    &\quad\quad\quad\quad\quad\quad\quad\quad\quad\left. - \frac{2c_{2}}{\sqrt{h}}\,T^{B}{}_{mn}\,\left({}^{\mathcal{A}}H^{ijklmn}\,\xi_{B} - 2\,h^{kn}\,h^{ln}\,\tilde{\xi}^{(ij)}\right)\right](x)\,\delta^{(3)}(\vec{x} - \vec{y})\\
    &\quad\quad\quad\quad\quad\quad\quad\quad\quad - \frac{2c_{2}}{\sqrt{h(x)}}\,h^{kn}(y)\,h^{lm}(y)\,\xi_{B}(y)\,\tilde{T}^{Bij}{}_{[m}(x)\partial^{(y)}_{n]}\,\delta^{(3)}(\vec{x} - \vec{y})\,,\\
\end{split}
\label{PB-algebra of AC and SC with upper indices}
\end{equation}
where we set
\begin{equation}
\begin{split}
    &{}^{\mathcal{A}}H^{ijklmn} := h^{lm}\,h^{i(k}\,h^{n)j} + h^{kn}\,h^{i(l}\,h^{m)j} + \frac{2}{3}\,h^{ij}\,h^{km}\,h^{ln}\,,\\
    &\tilde{T}^{Aij}{}_{m} := 2\,\theta^{A}{}_{a}\,h^{a(i}\,\delta^{j)}{}_{m} - \frac{2}{3}\,h^{ij}\,\theta^{A}{}_{m}\,,\\
    &\tilde{\xi}^{ij}{}_{A} := \frac{1}{12}\,h^{ja}\,\epsilon_{ABCD}\,\left(\epsilon^{imn}\,\theta^{B}{}_{a}\,\theta^{C}{}_{m}\,\theta^{D}{}_{n} + \epsilon^{lin}\,\theta^{B}{}_{l}\,\theta^{C}{}_{a}\,\theta^{D}{}_{n} + \epsilon^{lmi}\,\theta^{B}{}_{l}\,\theta^{C}{}_{m}\,\theta^{D}{}_{a}\right)\,.
\end{split}
\label{AH^ijklm tilde-T tilde-xi}
\end{equation}
In addition to this PB-algebra, the PB-algebra of ${}^{\mathcal{A}}\mathcal{C}^{ij}$ is given by
\begin{equation}
    \{{}^\mathcal{A}C^{ij}(t\,,\vec{x})\,,{}^\mathcal{A}C^{kl}(t\,,\vec{y})\} = \frac{2}{\sqrt{h}}\,\left(\delta^{j[l}\,{}^{\mathcal{A}}C^{k]i} + \delta^{i[k}\,{}^{\mathcal{A}}C^{l]j}\right)\,\delta^{(3)}(\vec{x}-\vec{y})\,,
\label{PB-algebra of AC}
\end{equation}
and the PB-algebra of ${}^{\mathcal{S}}\mathcal{C}^{ij}$ is already given by Eq.~(\ref{PB-algebra of SC and SC with upper indices}). 
Taking into account $c_{1} = c_{2} = 0$ in Type 7~\cite{Blixt:2018znp}, all the primary constraints ${}^{\mathcal{A}}\mathcal{C}^{ij}$ and ${}^{\mathcal{S}}\mathcal{C}^{ij}$ commute each other. 
However, as mentioned in Sec.~\ref{03:01Added}, the irregular property of Type 7 violates the linear independence of the diagonal components of ${}^{\mathcal{S}}\mathcal{C}^{ij} \propto {}^{\mathcal{S}}\pi^{ij}$; 
only two of three diagonal components are independent.
(See \hyperlink{thm01}{Regularity of NGR - (c)}.) 
In addition, differing from the case of Type 4, in Type 7, since all constraints are first-class, there is no room to give rise to further constraints without gauge-fixing procedures. 
Thus, ${}^{\mathcal{S}}\mathcal{C}^{ij} \approx 0$ provides only five independent first-class constraints, and Type 7 becomes a regular system subject to the five first-class constraints. 
The multiplier that corresponds to the excluded constraint is determined in relation to the other four multipliers when we impose a gauge-fixing condition. 
That is, the irregularity demands a reduction in the total number of first-class constraints to make the constraints functionally independent. 
The DoF of Type 7 is counted as follows: ${\rm DoF} = (16\times2 - 8\times2 - (3 + 5)\times2)/2 = 0\,$. 
Therefore, Type 7 is purely a topological system in the bulk spacetime. 

Unlike in the case of Type 4, the resulting constraint surface contains the original constraint surface due to the exclusion of one of the diagonal components of ${}^{\mathcal{S}}\mathcal{C}^{ij} \approx 0\,$; 
$\Gamma_{0} \subset \Gamma_{1}$ holds. 
This fact indicates that Type 7 is classified as Case B in Type I. 
Thus, in contrast to the case of Type 4, the result is valid on the entire constraint surface.
To investigate the possible physical DoFs in $\Gamma_{1}$, we consider to fix the gauge. 
Since a gauge-fixing condition can break the symmetry that consists of ${}^{\mathcal{S}}\mathcal{C}^{ij}$ and ${}^{\mathcal{A}}\mathcal{C}^{ij}$, extra DoFs can emerge, and the possible maximal number is three;
the minimal set of gauge conditions is to fix $\,{}^{\mathcal{A}}\mathcal{C}^{ij} \approx 0$ and $\,{}^{\mathcal{S}}\mathcal{C}^{ij} \approx 0$ by adding one constraint to each constraint, indicating that ${\rm DoF} = (16\times2 - 8\times2 - 4 - 6)/2 = 3\,$. 
More precisely, this gauge dependence of DoFs implies the appearance of bifurcation on $^\lnot\Gamma_{0} \cap\Gamma_{1}\,$. 
That is, there are three possible bifurcations; 
i), one gauge condition switches the constraint $\,{}^{\mathcal{A}}\mathcal{C}^{ij} \approx 0$ to second-class, resulting ${\rm DoF} = (16\times2 - 8\times2 - 4 - 5\times2)/2 = 1\,$; 
ii), one gauge condition switches the constraint $\,{}^{\mathcal{S}}\mathcal{C}^{ij} \approx 0$ to second-class, resulting ${\rm DoF} = (16\times2 - 8\times2 - 3\times2 - 6)/2 = 2\,$; 
iii), both cases i) and ii) occur at the same time, resulting ${\rm DoF} = (16\times2 - 8\times2 - 4 - 6)/2 = 3\,$. 
For each branch, the way of choosing a gauge is not unique but every gauge choice results the same DoFs; 
physics is independent of gauge choices in each branch. 
However, these DoFs are not necessarily those of Type 7 since it may occur from the subsurface $^\lnot\Gamma_{0} \cap\Gamma_{1}\,$; 
it seems that Type 7 of NGR may encounter the same issue in Chern-Simons theories with higher dimensions that is addressed in Ref.~\cite{Miskovic:2003ex}.

\subsection{\label{03:04}Specific sector in Type 9: DoF in Type 9}
The part of governing the proper symmetry in Type 9 of NGR is given by
\begin{equation}
     \tilde{\mathcal{H}}_{\rm Type\,9} = \sqrt{h}\,{}^{\mathcal{V}}\lambda_{i}(T^{*}\mathcal{Q})\,{}^{\mathcal{V}}\mathcal{C}^{i} + \sqrt{h}\,{}^{\mathcal{S}}\lambda_{ij}(T^{*}\mathcal{Q})\,{}^{\mathcal{S}}\mathcal{C}^{ij} + \sqrt{h}\,{}^{\mathcal{T}}\lambda(T^{*}\mathcal{Q})\,{}^{\mathcal{T}}\mathcal{C} + D_i[\pi_A{}^i(\alpha \xi^{A}+\beta^{j}\theta^A{}_j )]\,.
\label{specific part of H_type9}
\end{equation}
The DoF of this type is determined by the constraint structure of ${}^{\mathcal{V}}\mathcal{C}^{i} \approx 0$, ${}^{\mathcal{S}}\mathcal{C}^{ij} \approx 0$, and ${}^{\mathcal{T}}\mathcal{C} \approx 0$. 
The PB-algebra of ${}^{\mathcal{S}}\mathcal{C}^{ij}$ is already given in Eq.~(\ref{PB-algebra of SC and SC with upper indices}).
The PB-algebra of ${}^{\mathcal{T}}\mathcal{C}$ vanishes as a strong equality.
The PB-algebra of ${}^{\mathcal{V}}\mathcal{C}^{i}$ is calculated as follows:
\begin{equation}
    \{{}^{\mathcal{V}}C^{i}(t\,,\vec{x})\,,{}^{\mathcal{V}}C^{j}(t\,,\vec{y})\} = \frac{\eta^{00}}{\sqrt{h}}{}^{\mathcal{A}}C^{ij}\delta^{(3)}(\vec{x}-\vec{y})\,,
\label{PB-algebra of VC}
\end{equation}
where $\eta_{00}$ is the $(0\,,0)$-component of the Minkowski metric. 
The PB-algebra of ${}^{\mathcal{V}}\mathcal{C}^{i}$ and ${}^{\mathcal{S}}\mathcal{C}^{ij}$ is given by
\begin{equation}
\begin{split}
    &\{{}^{\mathcal{S}}C^{ij}(t\,,\vec{x})\,,{}^{\mathcal{V}}C^{k}(t\,,\vec{y})\} = \frac{1}{\sqrt{h}}\,\left[h^{k(i}\,{}^{\mathcal{V}}C^{j)} - \frac{1}{3}h^{ij}\,{}^{\mathcal{V}}C^{k} + \xi^{A}\,e_{A}{}^{k}\,{}^{\mathcal{S}}C^{ij}\right]\,\delta^{(3)}(\vec{x} - \vec{y})\\
    &\quad\quad\quad\quad\quad\quad\quad\quad\quad\quad - \frac{6c_{3}}{\sqrt{h}}\,\eta_{BC}\,\theta^{C}{}_{c}\,T^{B}{}_{ab}\,\left(h^{ka}\,{}^{\mathcal{V}}H^{ijbc} + \frac{1}{3}\,h^{bc}\,{}^{\mathcal{V}}H^{ijak}\right)\,\delta^{(3)}(\vec{x} - \vec{y})\\
    &\quad\quad\quad\quad\quad\quad\quad\quad\quad\quad - \frac{2c_{3}}{\sqrt{h}}\,h^{ka}(y)\,\theta_{B}{}^{b}(y)\,\tilde{T}^{Bij}{}_{[a}(x)\,\partial^{(y)}_{b]}\,\delta^{(3)}(\vec{x} - \vec{y})\,,
\end{split}
\label{PB-algebra of SC and VC with upper indices}
\end{equation}
where we set
\begin{equation}
    {}^{\mathcal{V}}H^{ijkl} := h^{i(k}\,h^{l)j} - \frac{1}{3}\,h^{ij}\,h^{kl}\,.
\label{VH^ijkl}
\end{equation}
In the case of Type 9, it set to be $c_{3} = 0$ in this PB-algebra~\cite{Blixt:2018znp}. 
The PB-algebra of ${}^{\mathcal{S}}\mathcal{C}^{ij}$ and ${}^{\mathcal{T}}\mathcal{C}$ is given Eq.~(\ref{PB-algebra of SC and TC with upper indices}).
The PB-algebra of ${}^{\mathcal{V}}\mathcal{C}^{i}$ and ${}^{\mathcal{T}}\mathcal{C}$ is calculated as follows:
\begin{equation}
    \{{}^{\mathcal{T}}C(t\,,\vec{x})\,,{}^{\mathcal{V}}C^{i}(t\,,\vec{x})\} = \frac{1}{\sqrt{h}}\,\left[\frac{1}{3}\,{}^{\mathcal{V}}C^{i} + \xi^{A}\,e_{A}{}^{i}\,{}^{\mathcal{T}}C\right]\,\delta^{(3)}(\vec{x} - \vec{y})
    - \frac{2}{\sqrt{h(x)}}\,\left( - 2\,c_{3}\,h^{ij}(y)\,\theta_{B}{}^{k}(y)\,\theta^{B}{}_{[j}(x)\,\partial^{(y)}_{k]}\,\delta^{(3)}(\vec{x} - \vec{y})\right)\,.
\label{PB-algebra of TC and VC}
\end{equation}
In the case of Type 9, it set to be $c_{3} = 0$ in this PB-algebra~\cite{Blixt:2018znp}. 
From these PB-algebras, in particular, Eq.~(\ref{PB-algebra of VC}) and Eq.~(\ref{PB-algebra of SC and TC with upper indices}), all the primary constraints, ${}^{\mathcal{V}}\mathcal{C}^{i} \approx 0$, ${}^{\mathcal{S}}\mathcal{C}^{ij} \approx 0$, and ${}^{\mathcal{T}}\mathcal{C} \approx 0$, are classified into second-class constraints. 
Therefore, all the multipliers are determined, and the Dirac procedure stops here. 
Since Type 9 is a regular system, (See \hyperlink{thm01}{Regularity of NGR - (a)},) the DoF can be counted out straightforwardly. 
The DoF of Type 9 is $(16\times2 - 8\times2 - (3 + 6 + 1))/2 = 3$. 

\subsection{\label{03:05}Interpretation of DoFs}
Both Type 4 and Type 9 have second-class constraints only. 
Type 7 is purely topological, and it shows no DoFs without fixing the gauge. 
Let us interpret the DoFs in former two types. 
From a generic perspective, as discussed in Ref.~\cite{Tomonari:2024ybs}, the existence of diffeomorphism symmetry of NGR suggests us to consider the possibility of breaking the local Lorentz Symmetry (LS) only~\cite{Mariz:2022oib}. 
Generic massive gravity never arises, which is caused by the violation of diffeomorphism symmetry, but breaking the local LS can also provide massive gravity~\cite{Rubakov:2004eb,Dubovsky:2004sg,Rubakov:2008nh}. 
However, as mentioned also in Ref.~\cite{Bahamonde:2024zkb,Tomonari:2024ybs}, these types do not contain any tensor propagating modes in advance. 

On the ghost DoFs, there is no Ostrogradski's ghost, since NGR is a quadratic theory. 
In fact, the Hamiltonians Eqs.~\eqref{specific part of H_type4} and~\eqref{specific part of H_type9} are bounded below. 
In Ref.~\cite{Bajardi:2024dru}, the Hamilton equations of NGR are derived, and we can also verify the absence of Ostrogradski's instability from these equations. 
At the linear perturbative level, one (pseudo-)scalar and two (pseudo-)vector modes can exist, and it is also possible for one ghost mode to arise due to the absence of propagating tensor modes. 
A more detailed nature of these DoFs requires that we consider a perturbative analysis around the FLRW spacetime~\cite{Cheng:1988zg,Izumi:2012qj}, but this is out of focus of the current work, leaving it for future investigations. 

In Type 4 and Type 9, these types do not contain two tensor DoFs that correspond to describing gravity~\cite{Bahamonde:2024zkb}. 
Furthermore, from a specific perspective, we can anticipate the upper bound of DoFs in the canonical analysis on each Type. 
In Type 4, since there are no generators of the local LS, the upper bound of the extra DoFs is six.
In terms of the Propagating Modes (PMs), all modes, one scalar, one pseudo-scalar, two vector, and two pseudo-vector modes, are possible to arise. 
In Type 9, since the generators ${}^{\mathcal{A}}C_{ij}$ are absent;
thus, the upper bound of the extra DoFs is three. 
In addition, the generators ${}^{\mathcal{V}}C_{i}$ could partially restore the local LS. 
That is, there is a possibility to revive one DoF relating to the local LS. 
Thus, in total, the upper bound of the extra DoFs is four. 
In terms of PMs, one scalar, one pseudo-scalar, and two pseudo-vector modes are possible to arise. 
These numbers are equal to or larger than the result of the canonical analysis in each Type, thus all the situations are consistent. 

\section{\label{04}Conclusions}
In this work, in Sec.~\ref{02}, we first reviewed NGR in the $SO(3)$-irreducible representation of canonical momenta. 
The fundamental ingredients required in performing the Dirac-Bergmann analysis were provided. 
Then, we proved a theorem, {\it i.e.,} \hyperlink{thm01}{Regularity of NGR}, which provides the classification of each type of NGR into either regular or irregular, and then we performed the analysis focusing on Type 4, Type 7, and Type 9. 
Then, Type 4 and Type 7 were irregular systems, whereas Type 9 was regular system.
It was unveiled that the full DoFs of these types are, in this order, five, zero, and three, respectively. 
Each type has no bifurcation. 
Type 4 and Type 9 have six second-class constraint densities and ten second-class constraint densities, respectively. 
Type 7 has eight first-class constraint densities.
The result of the analysis combining the previous work~\cite{Tomonari:2024ybs} is summarized in Table~\ref{Table:fullDoF}.

First-class constraint density in the internal space plays a crucial role in formulating physically interesting gravitational theories since such constraint density forms a closed algebra in the Poisson bracket. 
This indicates that the theory has a local symmetry~\cite{Sugano:1986xb,Sugano:1989rq,Sugano:1991ke,Sugano:1991kd,Sugano:1991ir}, and this symmetry provides propagating physically interesting modes including those of gravitation. 
In this regard, there is still room for Type 3 and Type 5 to be physically interested in theory at the non-linear level. 
The constraint densities of Type 8 do not form a closed algebra in the Poisson bracket, but in the smeared variables, the algebra in turn is closed. 
In addition, when imposing a specific condition to the Lagrange multiplier, the theory acquires a local gauge symmetry.
In the previous work~\cite{Tomonari:2024ybs}, we define a constraint density with such properties as semi-first class constraint density. 
In this jargon, we should investigate in detail the existence of and the nature of physically interesting mode in such a theory that a set of constraint densities classified as semi-first class exists.
Type 7 has only first-class constraint densities and is fulfilled by the possible number of constraints we can impose, thus it is ruled out at least for the purpose of describing gravity. 
That is, Type 7 is purely a topological system in bulk spacetime.

For conclusions, Type 2, Type 3, Type 5, Type 6 (TEGR), and Type 8, on the one hand, are interesting models to describe gravity, but at least the linear perturbation theory around the Minkowski background all types might contain strongly coupled modes in some scale and we need to remedy it by considering screening mechanism~\cite{Bahamonde:2024zkb}. 
Type 1 and Type 5 cannot be stable~\cite{Bahamonde:2024zkb}. 
On the other hand, Type 4, Type 7, and Type 9 are not interesting to describe gravity for cosmological applications, but these types may still be interesting as pure dynamical systems in other physics. 
In particular, Type 7 is a pure topological system in the bulk spacetime, but in a boundary of spacetime it may give rise to some intriguing non-vanishing dynamical DoF~\cite{Gallardo:2010er}. 
In Refs.~\cite{Chandia:1998uf,Miskovic:2003ex}, the authors investigate that an irregular constraint vanishes around the constraint surface. 
This reduction of irregular constraints affect directly to count out the number of DoFs. 
In particular, as mentioned in Sec.~\ref{03:03}, Type 7 may suffer from this issue; 
the subsurface $^\lnot\Gamma_{0} \cap\Gamma_{1}\,$ is not contained by the original constraint surface of Type 7, implying that there may occur an extra DoF. 
It would be an intriguing investigation to consider a linear perturbation around some background spacetime to verify whether the false arises or not in Type 7. 
We leave it for future work.
\begin{table}[ht!]
    \centering
    \renewcommand{\arraystretch}{1.5}
    \begin{tabular}{ c || c | c | c }
	Theory & Conditions & $SO(3)$-irreducible primary constraints & Nonlinear DoF\\ \hline \hline
	Type 1 & $A_{I}\neq 0 \ \forall I\in \{ \mathcal{V},{\mathcal{A}},{\mathcal{S}},\mathcal{T} \}$ &  No constraint & 8\\ \hline
	Type 2 & $A_{\mathcal{V}}=0$ &  ${}^{\mathcal{V}}\mathcal{C}_{i}\approx0$ & 6\\ \hline
	Type 3 & $A_{\mathcal{A}}=0$ &   ${}^{\mathcal{A}}\mathcal{C}_{ij}\approx0$ & 5\\ \hline
	Type 4 & $A_{\mathcal{S}}=0$ &   ${}^{\mathcal{S}}\mathcal{C}_{ij}\approx0$ & 5 \\ \hline
	Type 5 & $A_{\mathcal{T}}=0$ &   ${}^{\mathcal{T}}\mathcal{C}\approx0$ & 7\\ \hline
	Type 6 & $A_{\mathcal{V}}=A_{\mathcal{A}}=0$ & ${}^{\mathcal{V}}\mathcal{C}_{i}={}^{\mathcal{A}}\mathcal{C}_{ij}\approx0$ & 2\\ \hline
	Type 7 & $A_{\mathcal{A}}=A_{\mathcal{S}}=0$ &  ${}^{\mathcal{A}}\mathcal{C}_{ij}={}^{\mathcal{S}}\mathcal{C}_{ij}\approx0$ &  0 (Topological in bulk spacetime) \\ \hline
	Type 8 & $A_{\mathcal{A}}=A_{\mathcal{T}}=0$ &  ${}^{\mathcal{A}}\mathcal{C}_{ij}={}^{\mathcal{T}}C\approx0$ & 6 (Generic) or 4 (Special)\\ \hline
	Type 9 & $A_{\mathcal{V}}=A_{\mathcal{S}}=A_{\mathcal{T}}=0$ &  ${}^{\mathcal{V}}\mathcal{C}_{i}={}^{\mathcal{S}}\mathcal{C}_{ij}={}^{\mathcal{T}}C\approx0$ & 3 \\ \hline
    \end{tabular}
    \caption{Nonlinear DoF of each Type of NGR in the $SO(3)$-irreducible decomposition of canonical momentum. ``Special'' denotes the case that occurs only under the satisfaction of a set of specific conditions on Lagrange multipliers, whereas ``Generic'' denotes the case without any conditions. Type 6 is TEGR. For details in Type 1, Type 2, Type 3, Type 5, Type 6, and Type 8, see Ref.~\cite{Tomonari:2024ybs}.}
    \label{Table:fullDoF}
\end{table}

For a concluding remark on the series of our work, let us mention a potential issue of ADM-foliation. 
Recently, the concept of foliation and related aspects were reconsidered~\cite{Blixt:2024aej}. 
In this work, the authors indicate that a theory which is described by the (co-)frame field such as MAG and the extension/modification of MAG, which, of course, contains TEGR and NGR, may encounter difficulties when foliating four-dimensional spacetime into the 3+1 decomposition. 
They propose a necessary and sufficient condition for the existence of foliation. 
(See Theorem 2.3 in the work~\cite{Blixt:2024aej}.)
Quantitatively, this theorem amounts to vanishing the vorticity of the observer.
If the local Lorentz symmetry, on one hand, holds then for any observer a {\it global} foliation always exists and it does not depend on an observer.
On the other hand, if the local Lorentz symmetry is broken then a {\it global} foliation generically depends on each observer. 
It may imply that the result of the Dirac-Bergmann analysis of each type of NGR gives rise to further bifurcation. 
However, all considerations in the work~\cite{Blixt:2024aej} assume a certain coordinate system, but mathematically speaking, the concept of foliation is formulated without assuming any atlas of charts of a spacetime manifold. 
That is, the existence of foliation depends only on the topological structure of the spacetime manifold.
In fact, the global hyperbolicity ensures the existence of foliation. 
(See Theorem 8.3.14 in Ref.~\cite{Wald:1984rg} and Proposition 6.6.8 in Ref.~\cite{Hawking:1973uf}. 
One would find that all proofs in these references do not need any atlas of charts and assume topological structures only, thus the theorem is also applicable to the metric-affine geometry.) 
This point is very important for the past works in MAG and its extended/modified theories, of course, not excepting the works presented in the current and previous paper~\cite{Tomonari:2024ybs}.
One way to circumventing this difficulty is to implement differential forms to calculate all ingredients~\cite{Okolow:2011np,Okolow:2011nq,Okolow:2013lwa,Okolow:2023iyl}. 
In our perspective, the bifurcation ascribes only to the symmetry broken~\cite{Blagojevic:2020dyq,Blixt:2020ekl,Tomonari:2023wcs} but not to the coordinate choice, although. 
We leave the investigation of clarifying this point for future work.

\begin{acknowledgments}
KT would like to thank Daniel Blixt for insightful and fruitful discussions and the cosmology theory group in Institute of Science Tokyo for supporting my work, in particular, professor Teruaki Suyama. 
\end{acknowledgments}

\section*{Declarations}
\section*{Data availability}
Data sharing not applicable to this article as no datasets were generated or analysed during the current study.
\section*{Conflicts of interests}
The author have no competing interests to declare that are relevant to the content of this article.

\appendix
\section{\label{App:01}Complemental PB-algebras}
In addition to the PB-algebras given in Ref.~\cite{Tomonari:2024ybs}, we used the following PB-algebras in this work.

In Type 4, to complete the analysis, we need the following PB-algebras:
\begin{equation}
\begin{split}
    &\{{}^{\mathcal{S}}\pi^{ij}(t\,,\vec{x})\,,{}^{\mathcal{S}}\pi^{kl}(t\,,\vec{y})\} = \left(h^{ik}\,{}^{\mathcal{A}}\pi^{jl} + h^{il}\,{}^{\mathcal{A}}\pi^{jk} + h^{jk}\,{}^{\mathcal{A}}\pi^{il} + h^{jl}\,{}^{\mathcal{A}}\pi^{ik}\right)\delta^{(3)}(\vec{x} - \vec{y})\,,\\
    &\{\frac{1}{\sqrt{h(t\,,\vec{x})}}\,,{}^{\mathcal{S}}\pi^{ij}(t\,,\vec{y})\} = 0\,.    
\end{split}
\label{}
\end{equation}

In Type 7, to complete the analysis, we need the following PB-algebras:
\begin{equation}
\begin{split}
    &\{{}^{\mathcal{S}}\pi^{ij}(t\,,\vec{x})\,,{}^{\mathcal{A}}\pi^{kl}(t\,,\vec{y})\} = \left[\frac{1}{2}\,\left(h^{ik}\,{}^{\mathcal{A}}\pi^{jl} - h^{il}\,{}^{\mathcal{A}}\pi^{jk} + h^{jk}\,{}^{\mathcal{A}}\pi^{il} - h^{jl}\,{}^{\mathcal{A}}\pi^{ik}\right) + \frac{2}{3}\,h^{ij}\,{}^{\mathcal{A}}\pi^{kl}\right]\,\delta^{(3)}(\vec{x}-\vec{y})\,,\\
    &\{{}^{\mathcal{S}}\pi^{ij}(t\,,\vec{x})\,,h_{kl}(t\,,\vec{y})\} = - \left[2\,\delta^{i}_{(k}\,\delta^{j}_{l)} - \frac{2}{3}\,h^{ij}\,h_{kl}\right]\,\delta^{(3)}(\vec{x} - \vec{y})\,,\\
    &\{{}^{\mathcal{S}}\pi^{ij}(t\,,\vec{x})\,,h^{kl}(t\,,\vec{y})\} = \left[2\,h^{i(k}\,h^{l)j} - \frac{2}{3}\,h^{ij}\,h^{kl}\right]\,\delta^{(3)}(\vec{x} - \vec{y})\,,\\
    &\{{}^{\mathcal{S}}\pi^{ij}(t\,,\vec{x})\,,T^{A}{}_{kl}(t\,,\vec{y})\} = \left[2\,\theta^{A}{}_{a}\,h^{a(i}\,\delta^{j)}{}_{[k} - \frac{2}{3}\,h^{ij}\,\theta^{A}{}_{[k}\right](x)\,\partial^{(y)}_{l]}\,\delta^{(3)}(\vec{x} - \vec{y})\,,\\
     &\{{}^{\mathcal{S}}\pi^{ij}(t\,,\vec{x})\,,\xi_{A}(t\,,\vec{y})\} = \left[2\,\tilde{\xi}^{(ij)}{}_{A} - h^{ij}\,\xi_{A}\right]\,\delta^{(3)}(\vec{x} - \vec{y})\,.
\end{split}
\label{}
\end{equation}

In Type 9, to complete the analysis, we need the following PB-algebras:
\begin{equation}
\begin{split}
    &\{{}^{\mathcal{T}}\pi(t\,,\vec{x})\,,{}^{\mathcal{V}}\pi^{i}(t\,,\vec{y})\} = \frac{2}{3}\,{}^{\mathcal{V}}\pi^{i}\,\delta^{(3)}(\vec{x} - \vec{y})\,,\\
    &\{{}^{\mathcal{T}}\pi(t\,,\vec{x})\,,\theta_{A}{}^{i}(t\,,\vec{y})\} = \theta_{A}{}^{i}\,\delta^{(3)}(\vec{x} - \vec{y})\,,\\
    &\{{}^{\mathcal{S}}\pi^{ij}(t\,,\vec{x})\,,{}^{\mathcal{T}}\pi(t\,,\vec{y})\} = - \frac{1}{3}\,\left(\,{}^{\mathcal{S}}\pi^{ij} + \theta^{A}{}_{k}\,\pi_{A}{}^{(i}\,h^{j)k}\right)\,\delta^{(3)}(\vec{x} - \vec{y})\,,\\
    &\{{}^{\mathcal{S}}\pi^{ij}(t\,,\vec{x})\,,{}^{\mathcal{V}}\pi^{k}(t\,,\vec{y})\} = \left(h^{k(i}\,{}^{\mathcal{V}}\pi^{j)} - \frac{1}{3}\,h^{ij}\,{}^{\mathcal{V}}\pi^{k}\right)\,\delta^{(3)}(\vec{x} - \vec{y})\,,\\
    &\{\frac{1}{\sqrt{h(t\,,\vec{x})}}\,,{}^{\mathcal{V}}\pi^{i}(t\,,\vec{y})\} = \frac{1}{\sqrt{h}}\,\xi^{A}\,e_{A}{}^{i}\,\delta^{(3)}(\vec{x} - \vec{y})\,,\\
    &\{{}^{\mathcal{S}}\pi^{ij}(t\,,\vec{x})\,,\theta_{A}{}^{k}(t\,,\vec{y})\} = 3\,\eta_{AB}\,\theta^{B}{}_{l}\,{}^{\mathcal{V}}H^{ijkl}\,\delta^{(3)}(\vec{x} - \vec{y})\,.
\end{split}
\label{}
\end{equation}

\bibliographystyle{utphys}
\bibliography{Bibliography2.bib}

\providecommand{\href}[2]{#2}\begingroup\raggedright\begin{thebibliography}{10}

\bibitem{Planck:2018vyg}
{\bf Planck} Collaboration, N.~Aghanim {\em et al.}, ``{Planck 2018 results.
  VI. Cosmological parameters},''
  \href{http://dx.doi.org/10.1051/0004-6361/201833910}{{\em Astron. Astrophys.}
  {\bf 641} (2020)  A6}, \href{http://arxiv.org/abs/1807.06209}{{\tt
  arXiv:1807.06209 [astro-ph.CO]}}. [Erratum: Astron.Astrophys. 652, C4
  (2021)].

\bibitem{Vazquez:2018qdg}
J.~A. V\'azquez, L.~E. Padilla, and T.~Matos, ``{Inflationary cosmology: from
  theory to observations},''
  \href{http://dx.doi.org/10.31349/RevMexFisE.17.73}{{\em Rev. Mex. Fis. E}
  {\bf 17} (2020) no.~1, 73--91}, \href{http://arxiv.org/abs/1810.09934}{{\tt
  arXiv:1810.09934 [astro-ph.CO]}}.

\bibitem{Tsujikawa:2003jp}
S.~Tsujikawa, ``{Introductory review of cosmic inflation},'' in {\em {2nd Tah
  Poe School on Cosmology}: {Modern Cosmology}}.
\newblock 4, 2003.
\newblock \href{http://arxiv.org/abs/hep-ph/0304257}{{\tt
  arXiv:hep-ph/0304257}}.

\bibitem{Freese:2008cz}
K.~Freese, ``{Review of Observational Evidence for Dark Matter in the Universe
  and in upcoming searches for Dark Stars},''
  \href{http://dx.doi.org/10.1051/eas/0936016}{{\em EAS Publ. Ser.} {\bf 36}
  (2009)  113--126}, \href{http://arxiv.org/abs/0812.4005}{{\tt arXiv:0812.4005
  [astro-ph]}}.

\bibitem{Billard:2021uyg}
J.~Billard {\em et al.}, ``{Direct detection of dark matter\textemdash{}APPEC
  committee report*},'' \href{http://dx.doi.org/10.1088/1361-6633/ac5754}{{\em
  Rept. Prog. Phys.} {\bf 85} (2022) no.~5, 056201},
  \href{http://arxiv.org/abs/2104.07634}{{\tt arXiv:2104.07634 [hep-ex]}}.

\bibitem{SupernovaSearchTeam:1998fmf}
{\bf Supernova Search Team} Collaboration, A.~G. Riess {\em et al.},
  ``{Observational evidence from supernovae for an accelerating universe and a
  cosmological constant},'' \href{http://dx.doi.org/10.1086/300499}{{\em
  Astron. J.} {\bf 116} (1998)  1009--1038},
  \href{http://arxiv.org/abs/astro-ph/9805201}{{\tt arXiv:astro-ph/9805201}}.

\bibitem{SupernovaCosmologyProject:1998vns}
{\bf Supernova Cosmology Project} Collaboration, S.~Perlmutter {\em et al.},
  ``{Measurements of $\Omega$ and $\Lambda$ from 42 High Redshift
  Supernovae},'' \href{http://dx.doi.org/10.1086/307221}{{\em Astrophys. J.}
  {\bf 517} (1999)  565--586},
  \href{http://arxiv.org/abs/astro-ph/9812133}{{\tt arXiv:astro-ph/9812133}}.

\bibitem{H0LiCOW:2019pvv}
{\bf H0LiCOW} Collaboration, K.~C. Wong {\em et al.}, ``{H0LiCOW \textendash{}
  XIII. A 2.4 per cent measurement of H0 from lensed quasars:
  5.3\ensuremath{\sigma} tension between early- and late-Universe probes},''
  \href{http://dx.doi.org/10.1093/mnras/stz3094}{{\em Mon. Not. Roy. Astron.
  Soc.} {\bf 498} (2020) no.~1, 1420--1439},
  \href{http://arxiv.org/abs/1907.04869}{{\tt arXiv:1907.04869 [astro-ph.CO]}}.

\bibitem{Riess:2019cxk}
A.~G. Riess, S.~Casertano, W.~Yuan, L.~M. Macri, and D.~Scolnic, ``{Large
  Magellanic Cloud Cepheid Standards Provide a 1\% Foundation for the
  Determination of the Hubble Constant and Stronger Evidence for Physics beyond
  $\Lambda$CDM},'' \href{http://dx.doi.org/10.3847/1538-4357/ab1422}{{\em
  Astrophys. J.} {\bf 876} (2019) no.~1, 85},
  \href{http://arxiv.org/abs/1903.07603}{{\tt arXiv:1903.07603 [astro-ph.CO]}}.

\bibitem{Schoneberg:2022ggi}
N.~Sch\"oneberg, L.~Verde, H.~Gil-Mar\'\i{}n, and S.~Brieden, ``{BAO+BBN
  revisited \textemdash{} growing the Hubble tension with a 0.7 km/s/Mpc
  constraint},'' \href{http://dx.doi.org/10.1088/1475-7516/2022/11/039}{{\em
  JCAP} {\bf 11} (2022)  039}, \href{http://arxiv.org/abs/2209.14330}{{\tt
  arXiv:2209.14330 [astro-ph.CO]}}.

\bibitem{ACT:2023kun}
{\bf ACT} Collaboration, M.~S. Madhavacheril {\em et al.}, ``{The Atacama
  Cosmology Telescope: DR6 Gravitational Lensing Map and Cosmological
  Parameters},'' \href{http://dx.doi.org/10.3847/1538-4357/acff5f}{{\em
  Astrophys. J.} {\bf 962} (2024) no.~2, 113},
  \href{http://arxiv.org/abs/2304.05203}{{\tt arXiv:2304.05203 [astro-ph.CO]}}.

\bibitem{Tomonari:2024ybs}
K.~Tomonari and D.~Blixt, ``{Degrees of freedom of new general relativity: Type
  2, type 3, type 5, and type 8},''
  \href{http://dx.doi.org/10.1103/4ggt-6nd4}{{\em Phys. Rev. D} {\bf 112}
  (2025) no.~8, 084052}, \href{http://arxiv.org/abs/2410.15056}{{\tt
  arXiv:2410.15056 [gr-qc]}}.

\bibitem{Utiyama:1956sy}
R.~Utiyama, ``{Invariant theoretical interpretation of interaction},''
  \href{http://dx.doi.org/10.1103/PhysRev.101.1597}{{\em Phys. Rev.} {\bf 101}
  (1956)  1597--1607}.

\bibitem{Kibble:1961ba}
T.~W.~B. Kibble, ``{Lorentz invariance and the gravitational field},''
  \href{http://dx.doi.org/10.1063/1.1703702}{{\em J. Math. Phys.} {\bf 2}
  (1961)  212--221}.

\bibitem{Ivanenko:1983fts}
D.~Ivanenko and G.~Sardanashvily, ``{The Gauge Treatment of Gravity},''
  \href{http://dx.doi.org/10.1016/0370-1573(83)90046-7}{{\em Phys. Rept.} {\bf
  94} (1983)  1--45}.

\bibitem{Hehl:1994ue}
F.~W. Hehl, J.~D. McCrea, E.~W. Mielke, and Y.~Ne'eman, ``{Metric affine gauge
  theory of gravity: Field equations, Noether identities, world spinors, and
  breaking of dilation invariance},''
  \href{http://dx.doi.org/10.1016/0370-1573(94)00111-F}{{\em Phys. Rept.} {\bf
  258} (1995)  1--171}, \href{http://arxiv.org/abs/gr-qc/9402012}{{\tt
  arXiv:gr-qc/9402012}}.

\bibitem{Tomonari:2023ars}
K.~Tomonari, ``A unified-description of curvature, torsion, and non-metricity
  of the metric-affine geometry with the m\"obius representation,''
  \href{http://dx.doi.org/https://doi.org/10.1142/S021988782450333X}{{\em To be
  published in IJGMMP} (2024)  }, \href{http://arxiv.org/abs/2312.11558}{{\tt
  arXiv:2312.11558 [gr-qc]}}.

\bibitem{Blagojevic:2002du}
M.~Blagojevic, \href{http://dx.doi.org/10.1201/9781420034264}{{\em {Gravitation
  and gauge symmetries}}}.
\newblock 2002.

\bibitem{Adak:2005cd}
M.~Adak, M.~Kalay, and O.~Sert, ``{Lagrange formulation of the symmetric
  teleparallel gravity},''
  \href{http://dx.doi.org/10.1142/S0218271806008474}{{\em Int. J. Mod. Phys. D}
  {\bf 15} (2006)  619--634}, \href{http://arxiv.org/abs/gr-qc/0505025}{{\tt
  arXiv:gr-qc/0505025}}.

\bibitem{Adak:2006rx}
M.~Adak, ``{The Symmetric teleparallel gravity},'' {\em Turk. J. Phys.} {\bf
  30} (2006)  379--390, \href{http://arxiv.org/abs/gr-qc/0611077}{{\tt
  arXiv:gr-qc/0611077}}.

\bibitem{Adak:2008gd}
M.~Adak, O.~Sert, M.~Kalay, and M.~Sari, ``{Symmetric Teleparallel Gravity:
  Some exact solutions and spinor couplings},''
  \href{http://dx.doi.org/10.1142/S0217751X13501674}{{\em Int. J. Mod. Phys. A}
  {\bf 28} (2013)  1350167}, \href{http://arxiv.org/abs/0810.2388}{{\tt
  arXiv:0810.2388 [gr-qc]}}.

\bibitem{Adak:2011ltj}
M.~Adak and C.~Pala, ``{A novel approach to autoparallels for the theories of
  symmetric teleparallel gravity},''
  \href{http://dx.doi.org/10.1088/1742-6596/2191/1/012017}{{\em J. Phys. Conf.
  Ser.} {\bf 2191} (2022) no.~1, 012017},
  \href{http://arxiv.org/abs/1102.1878}{{\tt arXiv:1102.1878
  [physics.gen-ph]}}.

\bibitem{Bahamonde:2021gfp}
S.~Bahamonde, K.~F. Dialektopoulos, C.~Escamilla-Rivera, G.~Farrugia, V.~Gakis,
  M.~Hendry, M.~Hohmann, J.~S. Levi, J.~Mifsud, and E.~D. Valentino,
  ``{Teleparallel gravity: from theory to cosmology},''
  \href{http://dx.doi.org/10.1088/1361-6633/ac9cef}{{\em Rept. Prog. Phys.}
  {\bf 86} (2023) no.~2, 026901}, \href{http://arxiv.org/abs/2106.13793}{{\tt
  arXiv:2106.13793 [gr-qc]}}.

\bibitem{Hehl:1976kj}
F.~W. Hehl, P.~Von Der~Heyde, G.~D. Kerlick, and J.~M. Nester, ``{General
  Relativity with Spin and Torsion: Foundations and Prospects},''
  \href{http://dx.doi.org/10.1103/RevModPhys.48.393}{{\em Rev. Mod. Phys.} {\bf
  48} (1976)  393--416}.

\bibitem{Hayashi:1981fx}
K.~Hayashi and T.~Shirafuji, ``{Gravity From Poincare Gauge Theory of the
  Fundamental Particles. 6. Scattering Amplitudes},''
  \href{http://dx.doi.org/10.1143/PTP.66.318}{{\em Prog. Theor. Phys.} {\bf 66}
  (1981)  318}.

\bibitem{Blagojevic:2003cg}
M.~Blagojevic, ``{Three lectures on Poincare gauge theory},'' {\em SFIN A} {\bf
  1} (2003)  147--172, \href{http://arxiv.org/abs/gr-qc/0302040}{{\tt
  arXiv:gr-qc/0302040}}.

\bibitem{Hehl:2023khc}
F.~W. Hehl, ``{Four Lectures on Poincar\'e Gauge Field Theory},'' in {\em
  {International School of Cosmology and Gravitation: Spin, Torsion, Rotation
  and Supergravity}}.
\newblock 2023.
\newblock \href{http://arxiv.org/abs/2303.05366}{{\tt arXiv:2303.05366
  [gr-qc]}}.

\bibitem{Hayashi:1979qx}
K.~Hayashi and T.~Shirafuji, ``{New general relativity.},''
  \href{http://dx.doi.org/10.1103/PhysRevD.19.3524}{{\em Phys. Rev. D} {\bf 19}
  (1979)  3524--3553}. [Addendum: Phys.Rev.D 24, 3312--3314 (1982)].

\bibitem{Blixt:2018znp}
D.~Blixt, M.~Hohmann, and C.~Pfeifer, ``{Hamiltonian and primary constraints of
  new general relativity},''
  \href{http://dx.doi.org/10.1103/PhysRevD.99.084025}{{\em Phys. Rev. D} {\bf
  99} (2019) no.~8, 084025}, \href{http://arxiv.org/abs/1811.11137}{{\tt
  arXiv:1811.11137 [gr-qc]}}.

\bibitem{Pati:2022nwi}
L.~Pati, D.~Blixt, and M.-J. Guzman, ``{Hamilton\textquoteright{}s equations in
  the covariant teleparallel equivalent of general relativity},''
  \href{http://dx.doi.org/10.1103/PhysRevD.107.044071}{{\em Phys. Rev. D} {\bf
  107} (2023) no.~4, 044071}, \href{http://arxiv.org/abs/2210.07971}{{\tt
  arXiv:2210.07971 [gr-qc]}}.

\bibitem{Gomes:2023tur}
D.~A. Gomes, J.~Beltr\'an~Jim\'enez, A.~J. Cano, and T.~S. Koivisto,
  ``{Pathological Character of Modifications to Coincident General Relativity:
  Cosmological Strong Coupling and Ghosts in f(Q) Theories},''
  \href{http://dx.doi.org/10.1103/PhysRevLett.132.141401}{{\em Phys. Rev.
  Lett.} {\bf 132} (2024) no.~14, 141401},
  \href{http://arxiv.org/abs/2311.04201}{{\tt arXiv:2311.04201 [gr-qc]}}.

\bibitem{Bahamonde:2024zkb}
S.~Bahamonde, A.~Hell, D.~Blixt, and K.~F. Dialektopoulos, ``{Revisiting
  stability in new general relativity},''
  \href{http://dx.doi.org/10.1103/PhysRevD.111.064080}{{\em Phys. Rev. D} {\bf
  111} (2025) no.~6, 064080}, \href{http://arxiv.org/abs/2404.02972}{{\tt
  arXiv:2404.02972 [gr-qc]}}.

\bibitem{Vainshtein:1972sx}
A.~I. Vainshtein, ``{To the problem of nonvanishing gravitation mass},''
  \href{http://dx.doi.org/10.1016/0370-2693(72)90147-5}{{\em Phys. Lett. B}
  {\bf 39} (1972)  393--394}.

\bibitem{Hell:2021oea}
A.~Hell, ``{The strong couplings of massive Yang-Mills theory},''
  \href{http://dx.doi.org/10.1007/JHEP03(2022)167}{{\em JHEP} {\bf 03} (2022)
  167}, \href{http://arxiv.org/abs/2111.00017}{{\tt arXiv:2111.00017
  [hep-th]}}.

\bibitem{Dirac:1950pj}
P.~A.~M. Dirac, ``{Generalized Hamiltonian dynamics},''
  \href{http://dx.doi.org/10.4153/CJM-1950-012-1}{{\em Can. J. Math.} {\bf 2}
  (1950)  129--148}.

\bibitem{Dirac:1958sq}
P.~A.~M. Dirac, ``{Generalized Hamiltonian dynamics},''
  \href{http://dx.doi.org/10.1098/rspa.1958.0141}{{\em Proc. Roy. Soc. Lond. A}
  {\bf 246} (1958)  326--332}.

\bibitem{Bergmann:1949zz}
P.~G. Bergmann, ``{Non-Linear Field Theories},''
  \href{http://dx.doi.org/10.1103/PhysRev.75.680}{{\em Phys. Rev.} {\bf 75}
  (1949)  680--685}.

\bibitem{BergmannBrunings1949}
P.~G. Bergmann and J.~H.~M. Brunings, ``Non-linear field theories {II}.
  {C}anonical equations and quantization,''
  \href{http://dx.doi.org/10.1103/RevModPhys.21.480}{{\em Rev.Mod.Phys.} {\bf
  21} (1949)  480}.

\bibitem{Bergmann1950}
P.~G. Bergmann, R.~Penfield, R.~Schiller, and H.~Zatzkis, ``The {H}amiltonian
  of the general theory of relativity with electromagnetic field,''
  \href{http://dx.doi.org/10.1103/PhysRev.80.81}{{\em Phys.Rev.} {\bf 80}
  (1950)  81}.

\bibitem{Anderson:1951ta}
J.~L. Anderson and P.~G. Bergmann, ``{Constraints in covariant field
  theories},'' \href{http://dx.doi.org/10.1103/PhysRev.83.1018}{{\em Phys.
  Rev.} {\bf 83} (1951)  1018--1025}.

\bibitem{Dirac:1958sc}
P.~A.~M. Dirac, ``{The Theory of gravitation in Hamiltonian form},''
  \href{http://dx.doi.org/10.1098/rspa.1958.0142}{{\em Proc. Roy. Soc. Lond. A}
  {\bf 246} (1958)  333--343}.

\bibitem{Miskovic:2003ex}
O.~Miskovic and J.~Zanelli, ``{Dynamical structure of irregular constrained
  systems},'' \href{http://dx.doi.org/10.1063/1.1601299}{{\em J. Math. Phys.}
  {\bf 44} (2003)  3876--3887}, \href{http://arxiv.org/abs/hep-th/0302033}{{\tt
  arXiv:hep-th/0302033}}.

\bibitem{Tomonari:2023wcs}
K.~Tomonari and S.~Bahamonde, ``{Dirac\textendash{}Bergmann analysis and
  degrees of freedom of coincident f(Q)-gravity},''
  \href{http://dx.doi.org/10.1140/epjc/s10052-024-12677-x}{{\em Eur. Phys. J.
  C} {\bf 84} (2024) no.~4, 349}, \href{http://arxiv.org/abs/2308.06469}{{\tt
  arXiv:2308.06469 [gr-qc]}}. [Erratum: Eur.Phys.J.C 84, 508 (2024)].

\bibitem{Tomonari:2023vgg}
K.~Tomonari, ``{On the well-posed variational principle in degenerate point
  particle systems using embeddings of the symplectic manifold},''
  \href{http://dx.doi.org/10.1093/ptep/ptad073}{{\em PTEP} {\bf 2023} (2023)
  no.~6, 063A05}, \href{http://arxiv.org/abs/2304.00877}{{\tt arXiv:2304.00877
  [math-ph]}}.

\bibitem{Mariz:2022oib}
T.~Mariz, J.~R. Nascimento, and A.~Petrov,
  \href{http://dx.doi.org/10.1007/978-3-031-20120-2}{{\em {Lorentz Symmetry
  Breaking\textemdash{}Classical and Quantum Aspects}}}.
\newblock SpringerBriefs in Physics. Springer, 1, 2023.
\newblock \href{http://arxiv.org/abs/2205.02594}{{\tt arXiv:2205.02594
  [hep-th]}}.

\bibitem{Rubakov:2004eb}
V.~A. Rubakov, ``{Lorentz-violating graviton masses: Getting around ghosts, low
  strong coupling scale and VDVZ discontinuity},''
  \href{http://arxiv.org/abs/hep-th/0407104}{{\tt arXiv:hep-th/0407104}}.

\bibitem{Dubovsky:2004sg}
S.~L. Dubovsky, ``{Phases of massive gravity},''
  \href{http://dx.doi.org/10.1088/1126-6708/2004/10/076}{{\em JHEP} {\bf 10}
  (2004)  076}, \href{http://arxiv.org/abs/hep-th/0409124}{{\tt
  arXiv:hep-th/0409124}}.

\bibitem{Rubakov:2008nh}
V.~A. Rubakov and P.~G. Tinyakov, ``{Infrared-modified gravities and massive
  gravitons},'' \href{http://dx.doi.org/10.1070/PU2008v051n08ABEH006600}{{\em
  Phys. Usp.} {\bf 51} (2008)  759--792},
  \href{http://arxiv.org/abs/0802.4379}{{\tt arXiv:0802.4379 [hep-th]}}.

\bibitem{Bajardi:2024dru}
F.~Bajardi, D.~Blixt, and S.~Capozziello, ``{Hamilton equations in f(T)
  teleparallel gravity and in new general relativity},''
  \href{http://dx.doi.org/10.1103/PhysRevD.111.084012}{{\em Phys. Rev. D} {\bf
  111} (2025) no.~8, 084012}, \href{http://arxiv.org/abs/2412.20592}{{\tt
  arXiv:2412.20592 [gr-qc]}}.

\bibitem{Cheng:1988zg}
W.-H. Cheng, D.-C. Chern, and J.~M. Nester, ``{Canonical Analysis of the One
  Parameter Teleparallel Theory},''
  \href{http://dx.doi.org/10.1103/PhysRevD.38.2656}{{\em Phys. Rev. D} {\bf 38}
  (1988)  2656--2658}.

\bibitem{Izumi:2012qj}
K.~Izumi and Y.~C. Ong, ``{Cosmological Perturbation in f(T) Gravity
  Revisited},'' \href{http://dx.doi.org/10.1088/1475-7516/2013/06/029}{{\em
  JCAP} {\bf 06} (2013)  029}, \href{http://arxiv.org/abs/1212.5774}{{\tt
  arXiv:1212.5774 [gr-qc]}}.

\bibitem{Sugano:1986xb}
R.~Sugano, Y.~Saito, and T.~Kimura, ``{Generator of Gauge Transformation in
  Phase Space and Velocity Phase Space},''
  \href{http://dx.doi.org/10.1143/PTP.76.283}{{\em Prog. Theor. Phys.} {\bf 76}
  (1986)  283}.

\bibitem{Sugano:1989rq}
R.~Sugano and T.~Kimura, ``{Gauge Transformations for Dynamical Systems With
  First and Second Class Constraints},''
  \href{http://dx.doi.org/10.1103/PhysRevD.41.1247}{{\em Phys. Rev. D} {\bf 41}
  (1990)  1247}.

\bibitem{Sugano:1991ke}
R.~Sugano and Y.~Kagraoka, ``{Extension to velocity dependent gauge
  transformations. 1: General form of the generator},''
  \href{http://dx.doi.org/10.1007/BF01559438}{{\em Z. Phys. C} {\bf 52} (1991)
  437--442}.

\bibitem{Sugano:1991kd}
R.~Sugano and Y.~Kagraoka, ``{Extension to velocity dependent gauge
  transformations. 2. Properties of velocity dependent gauge
  transformations},'' \href{http://dx.doi.org/10.1007/BF01559439}{{\em Z. Phys.
  C} {\bf 52} (1991)  443--448}.

\bibitem{Sugano:1991ir}
R.~Sugano, Y.~Kagraoka, and T.~Kimura, ``{On gauge transformations and gauge
  fixing conditions in constraint systems},''
  \href{http://dx.doi.org/10.1142/S0217751X92000041}{{\em Int. J. Mod. Phys. A}
  {\bf 7} (1992)  61--90}.

\bibitem{Gallardo:2010er}
A.~Gallardo and M.~Montesinos, ``{The Boundary field theory induced by the
  Chern-Simons theory},''
  \href{http://dx.doi.org/10.1088/1751-8113/44/13/135402}{{\em J. Phys. A} {\bf
  44} (2011)  135402}, \href{http://arxiv.org/abs/1008.4883}{{\tt
  arXiv:1008.4883 [hep-th]}}.

\bibitem{Chandia:1998uf}
O.~Chandia, R.~Troncoso, and J.~Zanelli, ``{Dynamical content of Chern-Simons
  supergravity},'' \href{http://dx.doi.org/10.1063/1.59659}{{\em AIP Conf.
  Proc.} {\bf 484} (1999) no.~1, 231--237},
  \href{http://arxiv.org/abs/hep-th/9903204}{{\tt arXiv:hep-th/9903204}}.

\bibitem{Blixt:2024aej}
D.~Blixt, A.~Jim\'enez~Cano, and A.~Wojnar, ``{Foliation-generating observers
  under Lorentz transformations},'' \href{http://arxiv.org/abs/2408.16513}{{\tt
  arXiv:2408.16513 [gr-qc]}}.

\bibitem{Wald:1984rg}
R.~M. Wald,
  \href{http://dx.doi.org/10.7208/chicago/9780226870373.001.0001}{{\em {General
  Relativity}}}.
\newblock Chicago Univ. Pr., Chicago, USA, 1984.

\bibitem{Hawking:1973uf}
S.~W. Hawking and G.~F.~R. Ellis, {\em {The Large Scale Structure of
  Space-Time}}.
\newblock Cambridge Monographs on Mathematical Physics. Cambridge University
  Press, 2, 2023.

\bibitem{Okolow:2011np}
A.~Okolow and J.~Swiezewski, ``{Hamiltonian formulation of a simple theory of
  the teleparallel geometry},''
  \href{http://dx.doi.org/10.1088/0264-9381/29/4/045008}{{\em Class. Quant.
  Grav.} {\bf 29} (2012)  045008}, \href{http://arxiv.org/abs/1111.5490}{{\tt
  arXiv:1111.5490 [math-ph]}}.

\bibitem{Okolow:2011nq}
A.~Okol\'ow, ``{ADM-like Hamiltonian formulation of gravity in the teleparallel
  geometry},'' \href{http://dx.doi.org/10.1007/s10714-013-1605-y}{{\em Gen.
  Rel. Grav.} {\bf 45} (2013)  2569--2610},
  \href{http://arxiv.org/abs/1111.5498}{{\tt arXiv:1111.5498 [gr-qc]}}.

\bibitem{Okolow:2013lwa}
A.~Okol\'ow, ``{ADM-like Hamiltonian formulation of gravity in the teleparallel
  geometry: derivation of constraint algebra},''
  \href{http://dx.doi.org/10.1007/s10714-013-1636-4}{{\em Gen. Rel. Grav.} {\bf
  46} (2014)  1636}, \href{http://arxiv.org/abs/1309.4685}{{\tt arXiv:1309.4685
  [gr-qc]}}.

\bibitem{Okolow:2023iyl}
A.~Okolow, ``{Constraints of the Teleparallel Equivalent of General Relativity
  in a gauge},'' \href{http://arxiv.org/abs/2312.15991}{{\tt arXiv:2312.15991
  [gr-qc]}}.

\bibitem{Blagojevic:2020dyq}
M.~Blagojevi\'c and J.~M. Nester, ``{Local symmetries and physical degrees of
  freedom in $f(T)$ gravity: a Dirac Hamiltonian constraint analysis},''
  \href{http://dx.doi.org/10.1103/PhysRevD.102.064025}{{\em Phys. Rev. D} {\bf
  102} (2020) no.~6, 064025}, \href{http://arxiv.org/abs/2006.15303}{{\tt
  arXiv:2006.15303 [gr-qc]}}.

\bibitem{Blixt:2020ekl}
D.~Blixt, M.-J. Guzm\'an, M.~Hohmann, and C.~Pfeifer, ``{Review of the
  Hamiltonian analysis in teleparallel gravity},''
  \href{http://dx.doi.org/10.1142/S0219887821300051}{{\em Int. J. Geom. Meth.
  Mod. Phys.} {\bf 18} (2021) no.~supp01, 2130005},
  \href{http://arxiv.org/abs/2012.09180}{{\tt arXiv:2012.09180 [gr-qc]}}.

\end{thebibliography}\endgroup
\end{document}